\documentclass[amsfonts,amssymb,amsmath,showpacs]{revtex4}

\usepackage{bm}
\usepackage{graphicx}

\begin{document}

\title{Regular and singular solutions for charged dust distributions\\
   in the Einstein--Maxwell theory}

\author{Dubravko Horvat} \email{horvat@sfu.ca}
\affiliation{Department of Physics,
   Simon Fraser University, Burnaby, British Columbia, V5A\,1S6, Canada}
\affiliation{Department of Physics,
   Faculty of Electrical Engineering and Computing,
   University of Zagreb, Unska~3, HR-10\,000 Zagreb, Croatia}
\author{Sa{\v{s}}a Iliji{\'c}} \email{sasa.ilijic@fer.hr}
\affiliation{Department of Physics,
   Faculty of Electrical Engineering and Computing,
   University of Zagreb, Unska~3, HR-10\,000 Zagreb, Croatia}

\date{\today}

\begin{abstract}
Solutions for the static spherically symmetric
extremally charged dust in the Majumdar--Papapetrou system have been found.
For a certain amount of the allocated mass/charge,
the solutions have singularities of a type
which could render them physically unacceptable,
since the corresponding physically relevant quantities are singular as well.
These solutions, with a number of zero-nodes in the metric tensor,
are regularized through the $\delta$-shell formalism,
thus redefining the mass/charge distributions.
The bifurcating behaviour of regular solutions
found before is no longer present in these singular solutions,
but quantized-like behaviour in the total mass is observed.
Spectrum of regularized solutions
restores the equality of the Tolman--Whittaker and ADM mass,
as well the equality of the net charge and ADM mass,
which is the distinctive feature of Majumdar--Papapetrou systems.
\end{abstract}


\maketitle


\section{Introduction \label{sec:intro}}

In this paper we present a class of newly found singular, static,
spherically symmetric solutions to the Einstein--Maxwell equations
for the extremally charged dust (ECD) model of matter.
The solutions for the metric profile functions
show remarkable discrete properties
with respect to the amount of allocated mass.
These metric profile functions have nodes
that will be shown to correspond to naked singularities
and they are mimicking the behaviour of a quantum system
with regard to its higher and higher excitations.
Regularization procedure for the removal of the naked singularities
that are present in our solutions is investigated.
The regularization is accomplished by introducing
a thin shell ($\delta$-shell) of ECD into the system.
The regularized solutions retain their properties
with respect to the amount of the allocated mass,
as well as their asymptotic behaviour (charge and ADM mass).

We first briefly review the equations of structure
in the Majumdar-Papapetrou (MP) formalism \cite{Maj47,Papa47,LeZan05}.
In the next section we present the new solutions
with non-negative energy density of ECD.
These (singular) solutions exhibit zero crossings (nodes)
in the metric profile function.
In Sec.~\ref{sec:reg} we show
that introduction of a spherical $\delta$-shell of ECD
suffices to remove the spacetime singularity.
Conclusions are given in Sec.~\ref{sec:concl}.

The system is determined by the coupled Einstein--Maxwell equations,
  \begin{equation}
  R^{\mu}{}_{\nu} - \frac12 \, \delta^{\mu}_{\nu} \, R
  = 8\pi \, T^{\mu}{}_{\nu}
  \qquad \mathrm{and} \qquad
  \nabla_{\nu} F^{\nu\mu} = 4\pi\,j^{\mu} ,
  \label{eq:einmax}
  \end{equation}
where $T^{\mu}{}_{\nu}$ is the energy-momentum tensor,
$F^{\mu\nu}$ is the electromagnetic field strength tensor,
and $j^{\mu}$ is the electric charge 4-current.
Denoting $u^{\mu}$ the 4-velocity of the source
and $\sigma$ the charge density, we have $j^{\mu}=\sigma u^{\mu}$.
The energy-momentum tensor
due to the electrically charged (pressureless) dust,
$T^{\mu}{}_{\nu}$,
consists of the matter component
  \begin{equation}
  M^{\mu}{}_{\nu}=\rho\, u^{\mu} u_{\nu}
  \end{equation}
where $\rho$ is the energy density of the dust,
and the electromagnetic component
  \begin{equation}
  E^{\mu}{}_{\nu}
    = \frac{1}{4\pi} \Big( - F^{\mu\sigma} F_{\nu\sigma} +
  \frac{1}{4} \, \delta^{\mu}_{\nu} \, F^{\rho\sigma} F_{\rho\sigma} \Big)
  \label{eq:emunu}
  \end{equation}
due to field strength
$F_{\mu\nu} = \partial_{\mu} A_{\nu} - \partial_{\nu} A_{\mu}$.
As it was shown by Majumdar \cite{Maj47} and by Papapetrou \cite{Papa47},
if the energy density of the static dust equals the charge density,
$\rho=\pm\sigma$, one can write the line element
using the harmonic coordinates,
  \begin{equation}
  \mathrm{d} s^2 = U^{-2} \, \mathrm{d} t^2
    - U^{2} \, ( \mathrm{d} X^2 + \mathrm{d} Y^2 + \mathrm{d} Z^2 ) ,
  \label{eq:dsUXYZ}
  \end{equation}
where $U=U(X,Y,Z)$, and the Einstein--Maxwell equations (\ref{eq:einmax})
for the energy momentum tensor
$T^{\mu}{}_{\nu} = M^{\mu}{}_{\nu} + E^{\mu}{}_{\nu}$
reduce to a nonlinear version of the Poisson equation,
  \begin{equation}
  \nabla^2 U = - 4\pi \, \rho \, U^3
  \label{eq:einred}
  \end{equation}
($\nabla^2$ is the 3-dimensional flat space Laplacian operator).
Such matter is called the extremally charged,
or electrically counterpoised dust (ECD).
The 4-velocity of the static ECD is $u^{\mu}=U\,\delta^{\mu}_0$,
the charge 4-current is $j^{\mu}=\sigma\,U\,\delta^{\mu}_0$,
and the 4-potential has only the time component,
$A_{\mu}=\delta^0_{\mu}\Phi$,
where the electrostatic potential $\Phi$
is related to the metric profile function by $\Phi=\pm(U^{-1}-1)$.
For a formulation of the MP formalism in $d\ge4$ dimensional spacetimes,
as well as for a recent compilation of references
dealing with MP systems, see \cite{LeZan05}.

While one of the most remarkable features of the MP systems
is that they do not require any spatial symmetry,
spherically symmetric and asymptotically flat MP systems
have played a role in obtaining valuable insights
into the properties of astrophysically plausible systems,
and are sometimes called `Bonnor stars' \cite{BoWick75}.
In the spherical symmetry, the harmonic line element (\ref{eq:dsUXYZ}) is
  \begin{equation}
  \mathrm{d} s^2 = U(R)^{-2} \, \mathrm{d} t^2
    - U(R)^{2} \, (\mathrm{d} R^2 + R^2 \, \mathrm{d} \Omega^2) ,
  \label{eq:dsUR}
  \end{equation}
where $\mathrm{d}\Omega = \sin^2\vartheta \,
\mathrm{d}\varphi^2 + \mathrm{d}\vartheta^2$,
and the MP equation (\ref{eq:einred}) can be written
  \begin{equation}
  R^{-2}\,(R^2 U')' = - 4 \pi \, \rho \, U^3 ,
  \label{eq:mpR}
  \end{equation}
the prime denoting differentiation
with respect to the harmonic coordinate $R$.
The energy momentum tensor due to the electrostatic field (\ref{eq:emunu})
acquires the diagonal form:
  \begin{equation}
  E^{\mu}{}_{\nu} = \Phi'^2 \; \mathrm{diag} (+,+,-,-) / 8\pi.
  \end{equation}
The total energy density,
$\rho_{\mathrm{tot.}}=T^0{}_{0}=\rho+\Phi'^2/8\pi$,
involves the contribution from the ECD which is non-negative by assumption,
and the manifestly non-negative contribution from the electrostatic field.
The pressures are of purely electrostatic origin.
The transverse pressures, $p_{\mathrm{tr.}}
= - T^{\vartheta}{}_{\vartheta}
= - T^{\varphi}{}_{\varphi} = \Phi'^2/8\pi$,
are non-negative, while the radial pressure,
$p_{\mathrm{rad.}} = - T^R{}_{R} = - \Phi'^2/8\pi$,
is of equal strength, but of the opposite sign (tension).
These relations can be summarized as follows:
  \begin{equation}
  - p_{\mathrm{tr.}} = p_{\mathrm{rad.}} \le 0
  \le p_{\mathrm{tr.}} \le \rho_{\mathrm{tot.}} = p_{\mathrm{tr.}} + \rho.
  \end{equation}
The weak, null, strong, and the dominant energy conditions \cite{VissBook}
are all satisfied, as expected of classical matter and fields.

The Maxwell equation can be written
in terms of the Hodge-dual of the field strength tensor,
the 2-form ${}^*F_{\alpha\beta} = \epsilon_{\alpha\beta\mu\nu} F^{\mu\nu}$,
and the Hodge-dual of the charge 4-current,
the 3-form ${}^*j_{\alpha\beta\gamma}=\epsilon_{\alpha\beta\gamma\mu}j^{\mu}$,
as $\mathrm{d} {}^*F = 4\pi\,{}^*j$.
By the fundamental theorem of exterior calculus (Stokes's theorem),
one can use the Maxwell equation to express the charge $Q$
contained within a compact 3-region $\mathcal{R}$
as the integral of the field over the 2-boundary $\partial\mathcal{R}$,
$Q=\int_{\mathcal{R}} {}^*j
  = \frac1{4\pi} \int_{\mathcal{R}} \mathrm{d} {}^*F
  = \frac1{4\pi} \int_{\partial\mathcal{R}} {}^*F $.
In our context, we use the $t=\mathit{const.}$ slice
as the region $\mathcal{R}$, and the charge can be written as
  \begin{equation}
  Q = \int j^0 \, \sqrt{-g} \; \mathrm{d}^3 X
    = \mp \int \frac{ \nabla^2 U }{ 4\pi U^2 } \, \sqrt{-g} \; \mathrm{d}^3 X
    = \mp \int_{R_1}^{R_2} (R^2U')'\,\mathrm{d}R
    = \mp R^2 U' |_{R_1}^{R_2},
  \label{eq:Q}
  \end{equation}
where $R=R_1$ and $R=R_2$ are the radii of the inner and the outer
spherical 2-surfaces acting as boundaries of the 3-region $\mathcal{R}$
(we have used $j^0=\sigma\,U$ and $\sigma=\pm\rho=\mp\nabla^2U/4\pi U^3$).
In the regular center the electrical field vanishes
(i.e.\ $\Phi'=U'=0$ at $R=0$), and as $R_1\to0$,
the flux of the field through the inner boundary vanishes as well.
In fact, the inner boundary can be omitted,
and the total charge of the system is obtained
by letting $R_2\to\infty$, giving $Q=\mp\lim_{R\to\infty}R^2 U'$.

The MP geometries are static and therefore
they allow for the timelike Killing vector field $\xi^{\mu}$,
satisfying $\nabla_{(\mu}\xi_{\nu)}=0$,
orthogonal to the $t=\mathit{const.}$ slice.
Since the geometries are also asymptotically flat,
one normalizes the Killing vector
so that $\lim_{R\to\infty} \xi_\mu \xi^\mu = 1$.
The mass $M$ can be given by the Komar's formula \cite{Komar59}
as the integral of the 2-form $\alpha_{\alpha\beta}
= - \frac{1}{8\pi} \epsilon_{\alpha\beta\mu\nu} \nabla^{\mu}\xi^{\nu}$
over the 2-sphere $\mathcal{S}$ of sufficiently large radius,
$M=\int_{\mathcal{S}}\alpha$, which coincides with the ADM mass of the system.
However, by the Stokes's theorem, this integral can be expressed
as the integral of the 3-form $\mathrm{d}\alpha$
(the exterior derivative of $\alpha$) over the interior of the 2-sphere,
$\int_{\mathrm{int}(\mathcal{S})}\mathrm{d}\alpha$.
It can be shown \cite{WaldBook} that
$ (\mathrm{d}\alpha)_{\alpha\beta\gamma}
= - \frac1{4\pi}\epsilon_{\alpha\beta\gamma\mu}
    \nabla^{\nu}\nabla_{\nu} \xi^{\mu}
= \frac1{4\pi}\epsilon_{\alpha\beta\gamma\mu} R^{\mu}{}_{\nu} \xi^{\nu}
= \epsilon_{\alpha\beta\gamma\mu}
( 2T^{\mu}{}_{\nu} - \frac12 \delta^{\mu}_{\nu} T ) \xi^{\nu}$.
In the case of the static spacetimes, this reduces to the
Tolman--Whittaker expression \cite{Tolman30,Whittaker35},
which in our context integrates to
  \begin{equation}
  M = \int ( 2T^0{}_0 - T ) \sqrt{-g} \; \mathrm{d}^3 X
    = \int \left( \frac{U'^2}{4\pi U^4} - \frac{\nabla^2 U}{4\pi U^3} \right)
        \sqrt{-g} \; \mathrm{d}^3 X
    = - \frac{R^2 U'}{U}\Big|_{R_1}^{R_2},
  \label{eq:M}
  \end{equation}
where $R_1$ and $R_2$ are the radii
of the inner and the outer spherical surfaces (boundaries).
In a geometry with the regular center we can again omit the inner boundary,
and the mass of the system is obtained by letting $R_2\to\infty$,
i.e.\ $ M = - \lim_{R\to\infty} R^2 U' / U $.

Therefore, in asymptotically flat and nonsingular MP spacetimes we have
  \begin{equation}
  M = \pm Q = (\text{ADM mass}),
  \label{eq:MQADM}
  \end{equation}
where the mass $M$ and the charge $Q$
can be given by the integrals over the $t=\mathit{const.}$ slice,
and their values coincide with the value of the ADM mass
obtained at spatial infinity.
However, when singularities in the metric are present,
the `volume' integrals (\ref{eq:Q}) and (\ref{eq:M})
cannot be straightforwardly applied.
We deal with such solutions in the next Sections.

It is also important to give the relation among
the harmonic coordinates (\ref{eq:dsUR})
and the curvature coordinates with the line element
  \begin{equation}
  \mathrm{d} s^2 = B(r)\,\mathrm{d} t^2
     - A(r)\,\mathrm{d} r^2 - r^2 \, \mathrm{d} \Omega^2 ,
  \label{eq:dsABr}
  \end{equation}
where $r=UR$, and the metric components are related by
$A^{-1/2} = 1 + (\mathrm{d} U/\mathrm{d} R)\,R/U$, and $B=U^{-2}$.

In absence of ECD,
the general spherically symmetric solution of (\ref{eq:einred})
is $U(R)=k+m/R$, where $k$ and $m$ are integration constants.
To obtain asymptotically flat spacetime at infinity one sets $k=1$,
so in the spherically symmetric harmonic coordinates (\ref{eq:dsUR})
the metric profile function is
  \begin{equation}
  U(R) = 1 + m/R,
  \end{equation}
while in the curvature coordinates (\ref{eq:dsABr})
the metric profile functions read
  \begin{equation}
  B(r) = A^{-1}(r) = (1 - m/r)^2.
  \end{equation}
For $m>0$ this is the extremal Reissner--Nordstr\"om (ERN)
black hole spacetime of ADM mass $m$.
At the harmonic radius $R=0$ (corresponding to the curvature radius $r=m$),
there is the event horizon,
while at $R=-m$ ($r=0$), there is a point-like singularity.
In the case of negative ADM mass, $m<0$,
at $R=|m|$ (corresponding to $r=0$) there is a point-like singularity,
and since there is no event horizon in the spacetime,
this singularity is naked.

\section{The solutions \label{sec:sol}}

The MP solutions can be constructed
for the assumed distribution of the ECD
specified by the non-negative function $\rho(R)$.
If $\rho(R)$ vanishes as $R\to\infty$ more rapidly than $R^{-3}$,
the solutions are asymptotically flat at infinity.
This is expressed by the boundary conditions
  \begin{equation}
  \lim_{R\to\infty} U(R) = 1
  \quad \mathrm{and} \quad
  \lim_{R\to\infty} R^2 U'(R) = - m_{\infty} \; ,
  \label{eq:bc1}
  \end{equation}
where $m_{\infty}$ is the ADM mass of the system.

We will be looking for the solutions in which the
metric profile function $U$ is regular (finite) at $R=0$.
In the solutions in which $U$ has no nodes,
i.e.\ it is finite for all $R>0$,
the resulting spacetime will be free of singularities.
If, on the other hand,
$U(R)$ has one or more nodes at $R>0$,
then the outermost node corresponds to a naked pointlike singularity.
That the spacetime is singular can be seen
by observing that the Riemann quadratic invariant,
  \begin{multline}
  R^{\mu\nu\kappa\delta} R_{\mu\nu\kappa\delta}
  = \frac{4}{R^4 \, U^8} \, \Big[ 14 \, R^4 \, U'^4
    + 2 \, R^2 \, U \, {U'}^2 \big( 2 \, R \, U' - 5 \, R^2 \, U'' \big) \\
    + U^2 \, \big( 8 \, R^2 \, U'^2 + 4 \, R^3 \, U' \, U''
      + 3 \, R^4 \, {U''}^2 \big) \Big] ,
  \end{multline}
evidently diverges if for some $R_0>0$, $U(R_0)=0$.
The Ricci curvature scalar,
  \begin{equation}
  R = ( 4 \, U' + 2 \, R \, U'' ) / ( R U^3 ),
  \end{equation}
and the Weyl quadratic invariant,
  \begin{equation}
  C^{\mu\nu\kappa\delta} C_{\mu\nu\kappa\delta} =
  16 \, [ 3R \, {U'}^2 + U \, ( U' - R \, U'' ) ]^2 / ( 3R^2 \, U^8 ),
  \end{equation}
also diverge as $R\to R_0$.
The singularity is pointlike since the curvature radius $r$,
which is related to the radial coordinate $R$ by $r=R\,U(R)$,
tends to zero as $R\to R_0$.
The singularity is naked because there is no event horizon
to protect it from the outer space.

We will model the distribution of ECD with the exponential function
  \begin{equation}
  \rho(R) = \eta \, \rho_0 \, \exp[-(R/\tilde{R})^2],
  \end{equation}
where $\eta$ is the adjustable source strength factor, 
and $\rho_0$ and $\tilde{R}$ are constants, i.e.\ we will solve
  \begin{equation}
  R^{-2}\,(R^2 U')' = - 4\pi \, \eta \, \rho(R) \, U^3(R) .
  \label{eq:mpeta}
  \end{equation}
It was found before \cite{HIN05} that the regular solutions
do not exist for values of $\eta$
greater than some critical value $\eta_{\mathrm{c}}$,
while in the range $0<\eta<\eta_{\mathrm{c}}$,
the critical solution bifurcates
into two independent solutions with different ADM masses.
The `high mass bifurcation branch'
produces spacetimes that asymptotically coincide
with the external part of the ERN metric,
they allow arbitrarily large red-shifts \cite{BoWick75,Bonnor99},
and are known as `quasi black holes' \cite{HIN05,LeWein04}.

\begin{figure}
\begin{center} \includegraphics{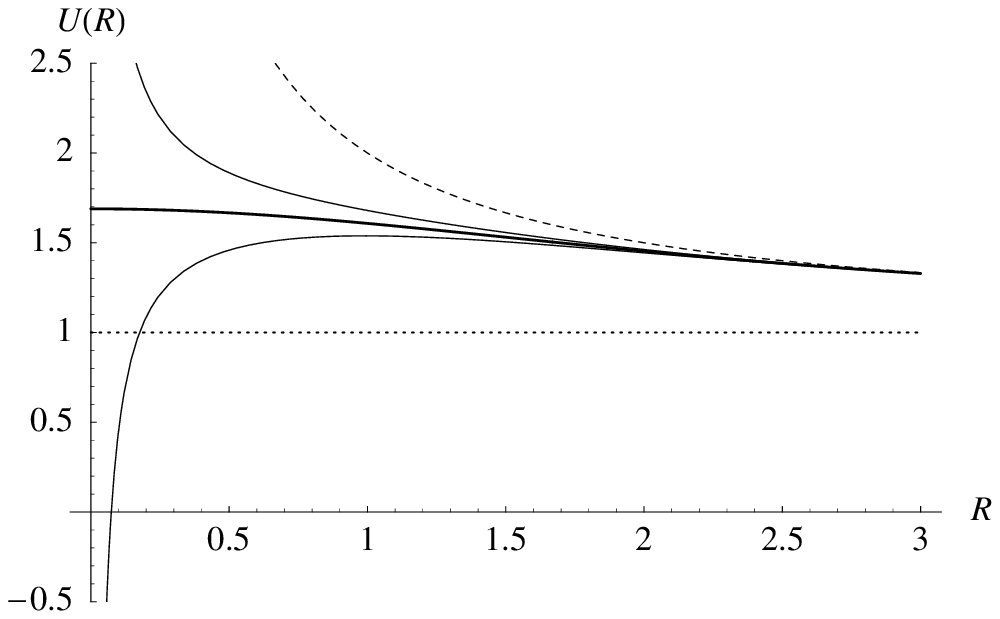} \end{center}
\begin{center} \includegraphics{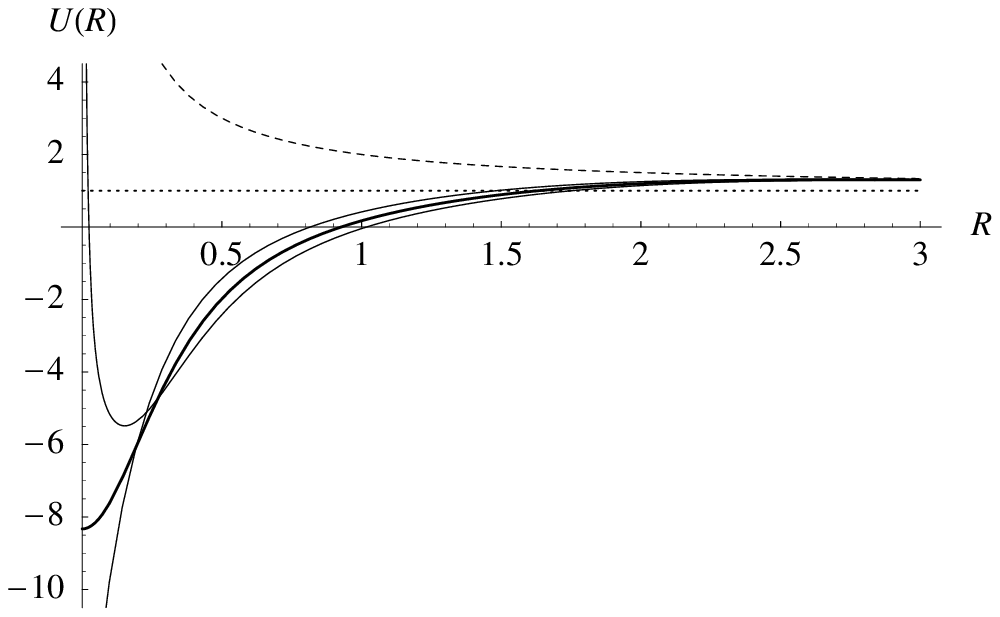} \end{center}
\caption{ \label{fig:1}
The metric profile function $U_0(R)$,
finite at $R=0$ with no nodes (thick line, upper plot),
and the function $U_1(R)$ finite at $R=0$ but with one node
(thick line, lower plot).
The metric profile functions with the same asymptotic mass,
corresponding to 20\% higher and 20\% lower source strength $\eta$
and diverging as $R\to 0$ are shown for both cases (thin lines).
Metric profile function for the ERN spacetime
is shown for comparison (dashed line).}
\end{figure}

\begin{figure}
\begin{center} \includegraphics{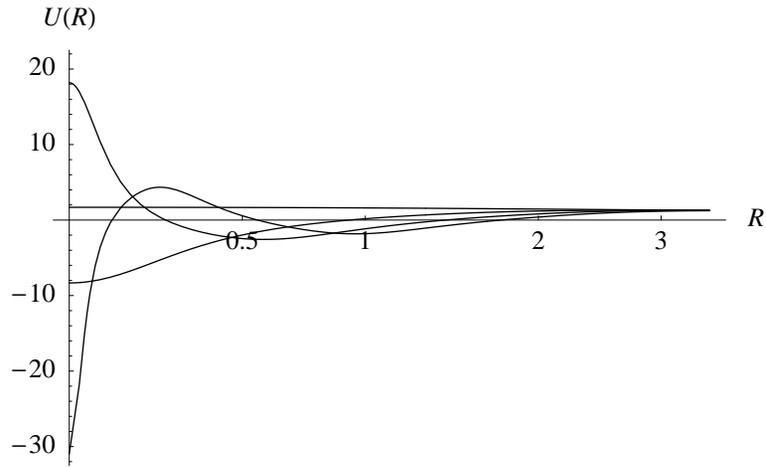} \end{center}
\caption{ \label{fig:2}
Metric profile functions $U_i(R)$ with $i=0,\dots,3$ nodes, finite at $R=0$,
are shown.  (Non-linear scaling is applied on the $R$ axis.)}
\end{figure}

We obtained the solutions to (\ref{eq:mpeta})
for the metric profile functions $U(R)$ with zero, one or more nodes,
subject to the boundary condition (\ref{eq:bc1})
with the fixed value of $m_{\infty}$ (ADM mass).
At $R=0$, we required that the metric profile function is finite
(see the above analysis), and we found that such solutions exist
only for a discrete set of values of the source strength factor $\eta$,
i.e.\ for discrete amounts of mass
allocated in the source term of (\ref{eq:mpeta}).
The `quantized' behaviour of the metric profile functions
with nodes is shown in Figs~\ref{fig:1} and \ref{fig:2}.

Starting from the solution $U_0(R)$
which is finite at $R=0$, $0<U_0(0) < \infty$,
and has no nodes, corresponding to source strength $\eta_0$,
a slight increase/decrease in the source strength,
$\eta_0 \to \eta_0\pm\delta \eta$,
makes $U(R)$ diverge towards $\mp\infty$ at $R=0$
(see Fig.~\ref{fig:1} upper plot).
If we increase the source strength further,
the divergence persists until a particular value of
the source strength $\eta_1$ is reached,
for which the solution $U_1(R)$ is again finite at $R=0$,
but now with a negative value, $-\infty<U_1(0)<0$,
thus $U_1(R)$ having one node (see Fig.~\ref{fig:1} lower plot).
In this way one can generate solutions with two,
three, and more nodes (see Fig.~\ref{fig:2}).

Since all these solutions obey the same boundary condition as $R\to\infty$,
they cannot be distinguished by a distant observer,
i.e.\ they all have the same ADM mass $m_{\infty}$.
For the regular solutions (no nodes),
the charge integral (\ref{eq:Q})
and Tolman--Whittaker mass integral (\ref{eq:M}),
integrated over the $t=\mathit{const.}$ slice,
obey the relation (\ref{eq:MQADM}), $M=\pm Q=m_{\infty}$.
In the case of the solutions with one or more nodes
in the metric profile function, the outermost being at $R=R_0$,
the integration of (\ref{eq:Q}) and (\ref{eq:M})
over the $t=\mathit{const.}$ slice from $R=R_0$ to $R=\infty$
does not give the expected value $m_{\infty}$.
This can be seen by looking at the nonvanishing contributions
of the fluxes through the infinitesimal inner boundary
(the 2-surface of radius $R\to R_0$, corresponding to $r\to 0$)
to the respective integrals.
In the case of the charge integral (\ref{eq:Q}),
this term contributes the finite value $Q_0=\pm R^2 U' |_{R_0}$,
thus rendering $Q= \pm m_{\infty} + Q_0$.
This can be understood as $-Q_0$
being the point charge located at the singularity.
In the mass integral (\ref{eq:M}),
the flux through the inner boundary diverges, leading to $M=\infty$.
In analogy with classical electrodynamics,
this can be understood as the diverging
electrostatic energy of the point charge.

\section{Regularization with the $\bm\delta$-shell \label{sec:reg}}

In view of the above results,
we now look for the possible modification of the ECD distribution
that can remove the pointlike naked singularity located at $R=R_0$
(corresponding to the outermost node of the metric profile function $U$)
while keeping the geometry outside of some radius $R=a>R_0$ unchanged.
In the same time we require that in the regularized solution
the charge integral (\ref{eq:Q})
and the Tolman--Whittaker mass integral (\ref{eq:M}),
when integrated over the $t=\mathit{const.}$ slice,
obey the relation (\ref{eq:MQADM}).

All the singular configurations considered so far
can be regularized by introducing a single spherical shell of ECD
with the regular geometry in the interior.
We will be using the $\delta$-shells with flat interior geometry,
expecting similar behaviour
also for the shells of finite thickness \cite{KZL05,MPfizbe}.
The flat geometry in the interior
(with constant metric profile function $U$)
is to be joined with the exterior geometry of our solutions,
given by the metric profile function $U(R)$ satisfying (\ref{eq:mpeta})
and with the outermost node at $R=R_0$,
along a spherical (timelike) hypersurface given by $R=a>R_0$.
Applying the Israel's junction condition formalism \cite{Israel66,VissBook},
we will obtain the components of the required energy-momentum
distribution on the junction hypersurface (shell).
On the both sides of the hypersurface we use the coordinates
$x^{\mu} = (t,R,\vartheta,\varphi)$,
while as the coordinates on the hypersurface
we naturally adopt $y^a = (t,\vartheta,\varphi)$.

The first junction condition requires
that the induced metric on the hypersurface
(the first fundamental form of the metric on the hypersurface),
given by $h_{ab} = g_{\mu\nu} \, e^{\mu}_a e^{\nu}_b$,
where $e^{\mu}_a = \partial x^{\mu} / \partial y^a$,
must be the same as induced by the metrics
on either side of the hypersurface,
i.e.\ the discontinuity of the induced metric
over the hypersurface must vanish.
In our context, this requires that the metric profile function $U$
is continuous over the hypersurface.
The regularized metric profile function $U_{\mathrm{reg}}$
can therefore be written
  \begin{equation}
  U_{\mathrm{reg}}(R) = U(a) +
  \big( U(R) - U(a) \big) \, \Theta(R-a),
  \end{equation}
where $\Theta(R)$ is the unit step function.

The second junction condition requires that the extrinsic curvature 
(the second fundamental form of the metric on the hypersurface),
given by $ K_{ab} = e^{\mu}_a e^{\nu}_b \, \nabla_{\nu} n_{\mu} $,
where $n_{\mu}=\delta_{\mu}^R U$ is the unit normal vector to the hypersurface,
must be the same as computed from the metric on either side
of the hypersurface.
Here, the nonzero components of the extrinsic curvature are
  \begin{equation}
  K^{0}{}_{0} = U'/U^2, \qquad 
  K^{\vartheta}{}_{\vartheta} = K^{\varphi}{}_{\varphi} = - (U/a+U')/U^2
  \end{equation}
where $U$ and $U'$ are evaluated at $R=a$.
Since in the flat interior we have $U'=0$,
while in the exterior we have $U'\ne0$ in general,
the second junction condition cannot be satisfied.
However, given that the first junction condition \emph{is} satisfied,
it is still possible to join the two geometries,
provided that we introduce (allow for) the appropriate
distribution of the energy-momentum on the hypersurface (shell).
The components of the energy-momentum tensor on the shell
are then given by the expression
  \begin{equation}
  S_{ab} = \frac1{8\pi} \Big( \big[\big[ K_{ab} \big]\big]
  - \big[\big[ K \big]\big] \, h_{ab} \Big),
  \end{equation}
where $K=K^a{}_a$ and the double brackets
denote the discontinuity of a quantity over the hypersurface.
In our context, for the surface energy density of the shell we obtain
  \begin{equation}
  \sigma=S^{0}{}_{0}=-{U'}/{4\pi U^2},
  \end{equation}
while the surface tension,
$\theta=S^{\vartheta}{}_{\vartheta}=S^{\varphi}{}_{\varphi}$,
vanishes identically.
From this one can see that the energy-momentum distribution
on the shell satisfies all standard energy conditions
if $\sigma$ is non-negative,
while all energy conditions are violated if $\sigma$ is negative.
Since in our solutions the ADM mass is positive,
the metric profile function $U$
is outwardly decreasing (\ref{eq:bc1}) as $R\to\infty$,
but since it is continous and nonnegative for $R>R_0$
it must have a maximum at some $R=R_{\max}>R_0$.
It follows that if the regularizing shell is introduced
at $R=a\in(R_0,R_{\max})$, i.e.\ where $U$ is outwardly increasing ($U'>0$),
all energy conditions are violated.
If it is introduced at $R=a>R_{\max}$
($U$ is outwardly decreasing, $U'<0$),
all energy conditions are satisfied.

In the regularized geometry,
the charge integral (\ref{eq:Q}) involves the contribution
due to the charge density on the shell
and due to the charge density of the
ECD distribution outside of the regularizing shell.
The resulting (regularized) charge density can be written,
  \begin{equation}
  j^0_{\mathrm{reg}} =
  \sigma \, \delta(R-a) + \rho \, \Theta(R-a) =
  - \frac{U'}{4\pi \, U^2} \; \delta(R-a)
  - \frac{\nabla^2 U}{4\pi\,U^3} \; \Theta(R-a)
  \end{equation}
and when integrated over the $t=\mathit{const.}$ slice
the expected value $m_{\infty}$ is obtained.
The Tolman--Whittaker mass integral (\ref{eq:M})
involves the expression
  \begin{equation}
  \left( 2 T^0{}_0 - T^{\mu}{}_{\mu} \right)_{\mathrm{reg}}
  = - \frac{U'}{4\pi \, U^3} \, \delta(R-a)
  + \left( \frac{U'^2}{4\pi U^4} - \frac{\nabla^2 U}{4\pi \, U^3} \right) \,
  \Theta(R-a)
  \end{equation}
where the first term is due to the energy-momentum contribution of the shell
and the second term is due to the exterior part
of the original ECD distribution.
The integral of this quantity over the $t=\mathit{const.}$ slice
yields $m_{\infty}$, as it should be.
Thus, the relation (\ref{eq:MQADM}) holds in the regularized system.

\section{Discussion and conclusions \label{sec:concl}}

We obtained a spectrum of asymptotically
extremal Reissner--Nordstr\"om (ERN) solutions
in the Majumdar--Papapetrou (MP) system.
The solutions are induced by the assumed distribution
of the extremally charged dust (ECD),
they have positive ADM mass,
the metric profile function $U(R)$ is finite at the harmonic radius $R=0$,
but it can have one or more nodes for $R>0$.
The outermost node corresponds to the naked pointlike singularity.
Such solutions are shown to exist only for a discrete set of values
of a parameter $\eta$ related to the source strength (allocated ECD/mass).
Here, for instance, the solutions with zero to three nodes
have source strengths: $1.00$, $7.29$, $19.6$, $39.1$.
The situation resembles a quantum system
with fixed boundary conditions,
in which to the normalized set of eigenfunctions
there corresponds a set of discrete eigenvalues
with higher ones corresponding to higher excitations
(and increasing number of nodes).

For a distant observer, the ADM mass of the MP spacetime
equals the $(\pm)$ total electrical charge of the system.
In the same time, the mass of the regular system
can be given by the Tolman--Whittaker mass (\ref{eq:M}),
and the total charge by the integral of the charge density (\ref{eq:Q}).
The spectrum of solutions with nodes in the metric profile function,
due to the presence of the singularity,
does not satisfy the expected relation
among the different mass and charge parameters (\ref{eq:MQADM}).
Also, since the singularities are naked,
the solutions can be taken as physically unacceptable.
We conclude the discussion of these solutions by showing that
by surrounding the singularity with a spherical $\delta$-shell of ECD
the solutions can be regularized.
Interestingly, depending on the position of the $\delta$-shell,
the energy density of the ECD can have either a positive or a negative sign
(see also \cite{BoCo89,HerVar94}).
For solutions with positive ADM mass and the outermost node is at $R=R_0$,
at some $R=R_{\max}>R_0$ there is a maximum of $U(R)$.
If the $\delta$-shell is placed in the region $R\in(R_0,R_{\max})$
the energy density of the shell is negative,
while if it is placed in the region $R\in(R_{\max},\infty)$,
the energy density is positive.
In the special case where the shell is placed exactly at $R=R_{\max}$,
the required amount of ECD on the shell is zero,
but in all these cases the required relation between mass and charge
which one finds in the regular MP systems is restored.
In the similar context, $\delta$-shells were used to regularize
the multiple black hole ERN spacetime in \cite{GurHim05}.
However, $\delta$-shell regularization is not the only possible approach;
thick shells \cite{LeZan05,MPfizbe} could nonetheless be applied
to achieve the similar results.
Alternatively, instead of altering the ECD distribution
to replace the central singularity with a patch of the regular spacetime,
one could, along the lines of Ref.~\cite{Balasin94},
also consider using the distributional techniques
to construct the distributional energy-momentum tensor
that allows the curvature singularity to be included in the manifold.

\begin{acknowledgments}
We thank the anonymous referee for suggestions
that helped us to improve the manuscript.
We also thank for hospitality the Simon Fraser University (D.H.)
and the University of Vienna (S.I.),
where part of this work was carried out.
This work is supported by the Croatian Ministry of Science
under the project 036-0982930-3144.
\end{acknowledgments}


\end{document}